\documentclass[prc,aps,superscriptaddress,preprint]{revtex4}
\usepackage{graphicx}
\begin{document}
\draft
\title                                                                            
{Spin-polarized neutron matter at different orders of chiral  
effective field theory 
}

\author{F. Sammarruca}
\email{fsammarr@uidaho.edu}
\affiliation{Department of Physics, University of Idaho, Moscow, ID 83844, USA}

\author{R. Machleidt}
\affiliation{Department of Physics, University of Idaho, Moscow, ID 83844, USA}

\author{N. Kaiser}
\affiliation{Physik Department, Technische Universit{\" a}t M{\" u}nchen, D-85747 Garching, Germany
}

\date{\today} 
\begin{abstract}
Spin-polarized neutron matter is studied using chiral two- and three-body forces.  
We focus, in particular, 
on predictions of the energy per particle in ferromagnetic neutron matter at 
different orders of chiral effective field theory and for different choices of the resolution scale.
We discuss the convergence pattern of the predictions and their cutoff dependence. 
We explore to which extent fully polarized 
neutron matter behaves (nearly) like a free Fermi gas. 
We also consider the more general case of partial polarization in neutron matter as well as 
the presence of a small proton fraction. In other words, in our calculations, we vary both spin
and isospin asymmetries.  
Confirming the findings of other microscopic calculations performed with different 
approaches, we report no evidence for a transition to a polarized phase of neutron matter.

\end{abstract}
\maketitle

\section{Introduction} 
                                                                     
The equation of state (EoS) of highly neutron-rich matter 
is a topic of current interest                                                 
because of its many applications ranging from the physics of 
rare isotopes to the properties of neutron stars. In spite of recent and fast-growing effort, 
the density dependence of the symmetry energy, which plays a chief role for the understanding of those systems, is not sufficiently constrained and, at the same time, theoretical predictions 
show considerable model dependence.                                    

Polarization properties of neutron/nuclear matter have been studied
extensively with a variety of
theoretical methods [1-25], often with contradictory conclusions. 
In the study in Ref.~\cite{IY}, for instance, 
the possibility of phase transitions into spin ordered states
of symmetric nuclear matter was explored based on the Gogny interaction 
\cite{Gogny} and the Fermi liquid formalism. In that paper, the appearance of 
an antiferromagnetic state (with opposite spins for neutrons and protons)
was predicted, whereas the transition to a ferromagnetic state was 
not indicated. This is in contrast to predictions based on      
Skyrme forces~\cite{I03}.

The properties of polarized neutron matter (NM) have gathered much attention lately, in conjunction with the issue of  ferromagnetic instabilities together with 
the possibility of strong magnetic fields in the interior
of rotating neutron stars.                                                          
The presence of polarization would impact neutrino cross sections and luminosities, resulting into a very 
different scenario for neutron star cooling. 

There are also other, equally important, motivations to undertake studies of polarized matter.
In Ref.~\cite{Sam10}, for instance, we focussed on the spin degrees of freedom of symmetric nuclear matter (SNM), having in 
mind a terrestrial scenario as a possible ``laboratory". 
We payed particular attention to the spin-dependent {\it symmetry potential}, namely the gradient between               
the single-nucleon potentials for upward and downward polarized nucleons in SNM. 
The interest around this quantity arises because of its natural interpretation as a 
spin dependent nuclear optical potential, defined in perfect formal analogy to the Lane potential \cite{Lane} for the isospin degree
of freedom in isospin-asymmetric nuclear matter (IANM). 

Whether one is interested in rapidly rotating pulsars or, more conventional, laboratory nuclear physics, it is important to consider 
both spin and isospin asymmetries. First, neutron star matter contains a 
non-negligible proton fraction. Concerning laboratory nuclear physics, 
one way to access information related to the spin dependence of the nuclear interaction in nuclear matter
is the study of collective modes such as giant resonances. Because a spin unsaturated system is usually 
also isospin asymmetric, both degrees of freedom need to be taken into account. 
For those reasons, 
in previous calculations~\cite{Sam11}, we extended our predictions~\cite{Sam10,SK07} to include matter with different   
concentrations of neutrons and protons where each nucleon species can have definite spin polarization. 
Our framework was based on the Dirac-Brueckner-Hartree-Fock (DBHF) approach to nuclear matter together with a 
realistic meson-theoretic potential.                                                            
Our findings did not show evidence of 
a phase transition to a ferromagnetic (FM) or antiferromagnetic (AFM) state.                                       
This conclusion appears to be shared by predictions of all microscopic models, such as those based on conventional Brueckner-Hartree-Fock theory 
\cite{pol18}. On the other hand, calculations based on various parametrizations of Skyrme forces result in 
different conclusions. For instance, 
 with the {\it SLy4} and {\it SLy5} forces and the Fermi liquid 
formalism                                                                                          
  a phase transition  to the AFM state is predicted in asymmetric matter 
at a critical density equal to about 2-3 times normal density \cite{IY}.
Qualitative disagreement is also encountered with other non-microscopic approaches such as                     
relativistic Hartree-Fock models based on effective meson-nucleon Lagrangians. For instance, in Ref.~\cite{pol12} 
it was reported that the onset of 
a ferromagnetic transition in neutron matter, and its critical density, are crucially determined by the inclusion of isovector mesons and the 
nature of their couplings. 

The brief review given above summarizes many useful and valid calculations. However, the 
problem common to all of them, including microscopic approaches, is that it is
essentially impossible to estimate, in a statistically meaningful way, the uncertainties
associated with a particular prediction, or to quantify the error related to the approximations 
applied in a particular model. 

Effective field theories (EFT) have shown the way out of this problem. 
Chiral effective field theory is a 
low-energy realization  of QCD 
\cite{Wei68,Wein79} which fits unresolved  nuclear dynamics at short
distances to the properties of two- and  few-nucleon systems.       
Together with a power 
counting, chiral EFT provides a framework where two and few-nucleon forces are generated on an equal 
footing in a systematic and controlled hierarchy.

Estimates of theoretical uncertainties \cite{furnstahl15} for calculations of the 
equation of state of nuclear and neutron matter have largely focused on varying the low-energy constants 
and resolution scale at which nuclear dynamics are probed 
\cite{bogner05,hebeler11,gezerlis13,krueger,coraggio13,coraggio14}. In a recent work~\cite{Sam+15}, 
we layed the foundations for order-by-order calculations of nuclear 
many-body systems by presenting consistent NLO and N$^2$LO chiral nuclear 
forces whose relevant short-range three-nucleon forces (3NF) are fit to $A=3$ binding 
energies and the lifetime of the triton~\cite{Marc12}. We then assessed the accuracy with which 
infinite nuclear and neutron matter properties and the isospin asymmetry energy can be 
predicted from order-by-order calculations in chiral effective field theory. 
In this paper, we apply the same philosophy to study the equation of state of 
polarized neutron matter. 

Based on the literature mentioned above,                              
a phase transition to a polarized phase (at least up to normal densities) seems unlikely, 
although the validity of such conclusion must be assessed in the context of EFT errors.
Furthermore, polarized neutron matter is a very interesting system for several reasons.
Because of the large neutron-neutron scattering length, NM displays behaviors similar to those 
of a unitary Fermi gas. In fact, 
up to nearly normal density, (unpolarized) neutron matter is found to display the behavior of an                  
$S$-wave superfluid~\cite{Carls03,Carls12}. 
The possibility of simulating low-density NM with 
ultracold atoms near a Feshbach resonance~\cite{Bloch08} has also been discussed.
When the system is totally polarized, 
it has been observed to behave like a weakly interacting Fermi gas~\cite{krueg14}. 
Here, we wish to explore to which extent and up to which densities we are in agreement with such conclusions, and 
how this and other observations depend on the chiral order and the resolution scale. 

In comparison with 
the calculations of Ref.~\cite{krueg14} (where 3NFs and 4NFs up to N$^3$LO were included), our present work
contains the following novelties:                                                            
\begin{itemize}
\item We consider both cutoff dependence and truncation error for the purpose of uncertainty 
quantification of chiral EFT.
Although incomplete in the 3NF at N$^3$LO, our calculations are a substantial step in that direction.
We note, further, that the contribution from the 3NF at N$^3$LO was found to be very small in neutron 
matter for the potentials in our perview~\cite{krueger}, about -0.5 MeV at normal density. Here, we consider
neutron matter or highly neutron-rich matter.
\item For the first time, we present results for both spin and isospin asymmetries within the framework of chiral forces.
As discussed in Section~\ref{res}, these tools are necessary to assess, for instance, the sensitivity 
of the results (particularly, the potential onset of a phase transition) to the presence of a proton fraction. 
\end{itemize}

This paper is organized as follows: In the next section, we present the formal aspects of the 
self-consistent calculation of the energy per particle, in general, applicable to infinite matter with any degree of isospin and spin asymmetry.
We also describe our approach to two- and three-body chiral forces. 
We provide expressions for the in-medium effective three-body force suitable for the most general case
of different proton and neutron concentrations where each species can be polarized to a different degree.
To the best of our knowledge, this has not been reported before in the literature within the framework
of chiral forces. 
Results for polarized and partially polarized NM, as well as for polarized netron-rich matter in the presence of a small proton fraction, are discussed in Section {\bf III}.
Conclusions and future plans are summarized in Section {\bf IV}. 

\section{Formalism } 
\label{form}
\subsection{General aspects} 
\label{GA} 

In a spin-polarized and isospin asymmetric system with fixed total density, $\rho$,               
the partial densities of each species are               
\begin{equation}
\rho_n=\rho_{nu}+\rho_{nd}\; , \; \; \; 
\rho_p=\rho_{pu}+\rho_{pd}\;, \; \; \; 
\rho=\rho_{n}+\rho_{p} \; ,           
\label{rho} 
\end{equation}
where $u$ and $d$ refer to up and down spin-polarizations, respectively, of protons ($p$) or neutrons ($n$). 
The isospin and spin asymmetries, $\alpha$, $\beta_n$, and $\beta_p$,  are defined in a natural way: 
\begin{equation}
\alpha=\frac{\rho_{n}-\rho_{p}}{\rho} \;, \; \; \;
\beta_n=\frac{\rho_{nu}-\rho_{nd}}{\rho_n} \;, \; \; \; 
\beta_p=\frac{\rho_{pu}-\rho_{pd}}{\rho_p} \;. 
\label{alpbet} 
\end{equation}
The density of each individual component can be related to the total density by 
\begin{equation}
\rho_{nu}=(1 + \beta_n)(1 + \alpha){\rho \over 4} \;, \; \; 
\rho_{nd}=(1 - \beta_n)(1 + \alpha){\rho \over 4}\; ,\;  \; 
\rho_{pu}=(1 + \beta_p)(1 - \alpha){\rho \over 4}\; ,\; \; 
\rho_{pd}=(1 - \beta_p)(1 - \alpha){\rho \over 4}\; , 
\label{rhopnud}
\end{equation}
where each partial density is related to the corresponding Fermi momentum 
through $\rho_{\tau \sigma}$ =$ (k_F^{\tau \sigma})^3/(6\pi^2)$. 
The {\it average} Fermi momentum  and the total density are related in the usual way as 
$\rho= (2 k_F^3)/(3 \pi ^2)$. 

The single-particle potential of a nucleon in a particular $\tau \sigma$ state, $U_{\tau \sigma}$, is the solution of a
set of four coupled equations, 
\begin{equation}
U_{nu} = U_{nu,nu} + U_{nu,nd} + U_{nu,pu} + U_{nu,pd}       
\label{unu}
\end{equation} 
\begin{equation} 
U_{nd} = U_{nd,nu} + U_{nd,nd} + U_{nd,pu} + U_{nd,pd}        
\label{und}
\end{equation} 
\begin{equation} 
U_{pu} = U_{pu,nu} + U_{pu,nd} + U_{pu,pu} + U_{pu,pd}      
\label{upu}
\end{equation} 
\begin{equation} 
U_{pd} = U_{pd,nu} + U_{pd,nd} + U_{pd,pu} + U_{pd,pd}   \; , 
\label{upd}
\end{equation}
to be solved self-consistently along with the effective interaction, the $G$-matrix.                   
(The latter will be discussed in the next two subsections.)
In the above equations,                                                              
each $U_{\tau \sigma, \tau '\sigma'}$ term on the right-hand side contains the
appropriate (spin and isospin dependent) part of the interaction, $G_{\tau \sigma,
'\tau'\sigma'}$. More specifically,
\begin{equation}
U_{\tau \sigma,\tau' \sigma'}({\vec k}) =  \sum _{q\leq k_F^{\tau' \sigma
'}} <\tau \sigma,\tau'\sigma'|G({\vec k},{\vec q})|\tau \sigma,\tau'\sigma'>,
\label{ug}   
\end{equation}
where the summation indicates integration over the Fermi
seas of protons and neutrons with spin-up and spin-down, and                           
\begin{widetext}
\begin{eqnarray}
<\tau \sigma,\tau'\sigma'|G({\vec k},{\vec q})|\sigma \tau,\sigma '\tau'>&=&
\sum_{L,L',S,J,M,M_L,T} |<\frac{1}{2} \sigma;\frac{1}{2} \sigma '|S
(\sigma + \sigma ')>|^2
|<\frac{1}{2} \tau;\frac{1}{2} \tau '|T
(\tau + \tau ')>|^2 
\nonumber\\ &&  
\times<L M_L;S(\sigma + \sigma ')|JM>
<L' M_L;S(\sigma + \sigma ')|JM> \nonumber\\ &&
\times i^{L'-L} Y^{*}_{L',M_L}({\hat k_{rel}}) Y_{L,M_L}({\hat
k_{rel}}) <LSJ|G(k_{rel},K_{c.m.})|L'SJ> \; . 
\label{gmat} 
\end{eqnarray}
\end{widetext}
The $G$-matrix which appears in the formulas above is constructed from the two-nucleon                    
potential 
and the effective density-dependent 3NF as explained later. 

The need to separate the interaction by spin
components brings along angular dependence, with the result that the
single-particle potential depends also on the direction of the
momentum, although such dependence was found to be weak \cite{SK07}. The $G$-matrix equation is solved using
partial wave decomposition and the matrix elements are then summed
as in Eq.~(\ref{gmat}) to provide the new matrix elements in the
representation needed for Eq.~(\ref{ug}), namely with spin and isospin components explicitely projected out. Furthermore, the
scattering equation is solved using relative and center-of-mass
coordinates, $k_{rel}$ and $K_{c.m.}$, since the former is a natural coordinate for the evaluation of the nuclear potential. Those are then easily related
to the momenta of the two particles, $k$ and $q$, in order to
perform the integration indicated in Eq.~(\ref{ug}).  Notice that solving
the $G$-matrix equation requires knowledge of the single-particle
potential, which in turn requires knowledge of the effective interaction.
Hence, Eqs.(~\ref{unu}-\ref{upd}) together with the $G$-matrix equation constitute a
rather lengthy self-consistency problem,                                                  
the solution of which yields the single-nucleon
potentials in each $\tau \sigma$ channel. 

The kernel of the $G$-matrix equation contains 
the Pauli operator for scattering of two particles with
two different Fermi momenta, $k_F^{\tau \sigma}$ and 
 $k_F^{\tau' \sigma'}$, which is  
defined in analogy with the
 one for isospin-asymmetric matter~\cite{AS1},
\begin{equation}
Q_{\tau \sigma, \tau' \sigma'}(k,q,k_F^{\tau \sigma},k_F^{'\tau'\sigma'})=\left\{
\begin{array}{l l}
1 & \quad \mbox{if $p>k_F^{\tau \sigma}$ and  $q>k_F^{\tau' \sigma'}$}\\
0 & \quad \mbox{otherwise.}
\end{array}
\right.
\label{pauli} 
\end{equation}
The Pauli operator is expressed in terms of $k_{rel}$ and
$K_{c.m.}$ and angle-averaged in the usual way.
We then proceed with the calculation of the energy per nucleon in the particle-particle ladder approximation, namely the leading-order 
contribution in the hole-line expansion. (See Ref.~\cite{Sam+15} and references therein for a discussion of the 
uncertainty associated with this approximation.) 

Once a self-consistent solution for Eqs.~(\ref{unu}-\ref{upd}) has been obtained, the average potential energy for 
a given $\tau \sigma$ component can be calculated. 
A final average over all              
$\tau \sigma$ components provides, along with the kinetic energy $K_{\tau \sigma}$, the average energy per particle in spin-polarized
isospin-asymmetric nuclear matter. 
Specifically, 
\begin{equation}
\frac{E}{A} = \frac{1}{A} \sum _{\sigma =u,d}\sum_{\tau=n,p} \sum _{k\leq k_F^{\tau \sigma     
}} \Big (K_{\tau \sigma}(k) + \frac{1}{2} U_{\tau \sigma}(k) \Big ) \; ,         
\label{ea}
\end{equation}
where $E/A$ is a function of $\rho$, $\alpha$, $\beta_n$, and $\beta_p$, with $\alpha$=1 in the present case.
All calculations are conducted including values of the total angular momentum $J$ from 0 to 15.

\subsection{Chiral two-body potentials}                                                                  
\label{2nf}

\begin{table}
\centering
\begin{tabular}{|c||c|c|c|c|c|}
\hline
NLO & $\Lambda$ (MeV) & $n$& $c_1$ & $c_3$ & $c_4$ \\
\hline     
 &  450 & 2&       &       &       \\
 &  500 & 2&       &       &       \\
 &  600 & 2&       &       &       \\
\hline
\hline
N$^2$LO & $\Lambda$ (MeV) & $n$& $c_1$ & $c_3$ & $c_4$ \\
\hline     
 & 450 & 3& -0.81 & -3.40 & 3.40  \\
 & 500 & 3& -0.81 & -3.40 & 3.40  \\
 & 600 & 3& -0.81 & -3.40 & 3.40  \\
\hline
\hline
N$^3$LO & $\Lambda$ (MeV) & $n$& $c_1$ & $c_3$ & $c_4$ \\
\hline     
 & 450 & 3& -0.81 & -3.40 & 3.40  \\
 & 500 & 2& -0.81 & -3.20 & 5.40  \\
 & 600 & 2& -0.81 & -3.20 & 5.40  \\
\hline
\hline
\end{tabular}
\caption{Values of $n$ and low-energy constants of the dimension-two $\pi N$ Lagrangian, 
$c_{1,3,4}$, at each order and for each type of cutoff in the regulator function given in 
Eq.~(\ref{reg}). None of the $c_i$'s appears at NLO. The low-energy constants are given in 
units of GeV$^{-1}$.}
\label{tab1}
\end{table}

In this section we discuss in some detail the features of the 
nucleon-nucleon ($NN$) potentials we use for these calculations.

All low-momentum interactions are 
limited in calculations of the EoS to densities where the characteristic 
momentum scale (on the order of the Fermi momentum) is below the scale 
set by the momentum-space cutoff $\Lambda$
in the $NN$ potential regulating function, which for chiral $NN$ forces
typically has the form:
\begin{equation}
f(p',p) = \exp[-(p'/\Lambda)^{2n} - (p/\Lambda)^{2n}] \; ,
\label{reg}
\end{equation}
where $\Lambda \lesssim 500$\,MeV is associated with the onset of
favorable perturbative properties~\cite{coraggio13,coraggio14}. 

Although designed to reproduce similar $NN$ scattering phase shifts, $NN$ 
potentials with different regulator functions will yield different predictions in the 
nuclear many-body problem due to their different off-shell behavior.  On the 
other hand, appropriate re-adjustment of the low-energy constants that appear 
in the nuclear many-body forces is expected to reduce the dependence on the 
regulator function \cite{coraggio13}.

In the present investigation we consider $NN$ potentials at order $(q/\Lambda_\chi)^2$,
$(q/\Lambda_\chi)^3$ and $(q/\Lambda_\chi)^4$ in the chiral power counting, where 
$q$ denotes the small scale set by external nucleon momenta or the pion mass and 
$\Lambda_\chi$ is the chiral symmetry breaking scale. Chiral $NN$ potentials at NLO
and N$^2$LO, corresponding to $(q/\Lambda_\chi)^2$ and $(q/\Lambda_\chi)^3$, 
have been constructed previously in Ref.~\cite{NLO} for cutoffs ranging from $\Lambda$
= 450 MeV to about 800 MeV. With varying chiral order and cutoff scale, the low-energy 
constants in the two-nucleon sector are refitted to elastic $NN$ scattering 
phase shifts and properties of the deuteron. The low-energy constants $c_{1,3,4}$ 
associated with the $\pi \pi N N$ contact couplings of the ${\cal L}^{(2)}_{\pi N}$ chiral
Lagrangian are given in Table~\ref{tab1}.
We note that the $c_{i}$ can be extracted from $\pi N$ or $NN$ scattering data. The potentials we use here
\cite{EM03,ME11} follow the second path. At N$^2$LO, taking the range determined 
in analyses of elastic $\pi N$ scattering as a starting point, values were chosen to best reproduce $NN$ data at that order. 
At N$^3$LO, high-precision required a stronger adjustment of $c_4$ depending on the regulator 
function and cutoff. The fitting procedure is discussed in Ref.~\cite{ME11}, where it is noted that 
the larger value for $c_4$ has, overall, a very small impact but lowers the $^3F_2$ phase shift 
for a better agreement with the phase shift analysis.

In Ref.~\cite{NLO}, it was found that the two-body scattering phase shifts can be 
described well at NLO up to a laboratory energy of about 100 MeV, while the N$^2$LO 
potential fits the data up to 200 MeV. Interestingly, in the latter case the $\chi^2/$datum was 
found to be essentially cutoff independent for variations of $\Lambda$ between 450 
and approximately 800 MeV. Finally, we also use $NN$ potentials constructed 
at next-to-next-to-next-to-leading order (N$^3$LO) \cite{ME11,EM03}, with low-energy 
constants $c_{1,3,4}$ as displayed in Table~\ref{tab1}.                             

Although N$^2$LO calculations can achieve sufficient
accuracy in selected partial wave channels up to $E_{\rm lab} = 200$\,MeV, only the
N$^3$LO interactions achieve the level of high-precision potentials, characterized by a 
$\chi^2/$datum $\sim 1$.

At the two-body level, each time the chiral order is increased, the $NN$ contact terms 
and/or the two-pion-exchange contributions proportional to the low-energy constants
$c_{1,3,4}$ are refitted. We recall that at N$^2$LO no new $NN$ contact terms
are generated, and therefore improved cutoff independence in the $NN$ phase 
shifts~\cite{Sam+15} is due to changes in the 
two-pion-exchange contributions. At N$^2$LO, subleading $\pi \pi N N$ vertices 
enter into the chiral $NN$ potential. These terms encode the important physics of 
correlated two-pion-exchange and the excitation of intermediate $\Delta$(1232) 
isobar states. Therefore, at this order it is possible to obtain a realistic description of the 
$NN$ interaction at intermediate range, traditionally generated through the exchange of 
a fictitious $\sigma$ meson of medium mass. At N$^3$LO in the chiral power 
counting, 15 additional $NN$ contact terms (bringing the total number to 24 at N$^3$LO) 
result in a much improved description of $NN$ scattering phase shifts.

\subsection{The three-nucleon force}                                                                  
\label{3nf}

\begin{figure}[!t] 
\centering
\includegraphics{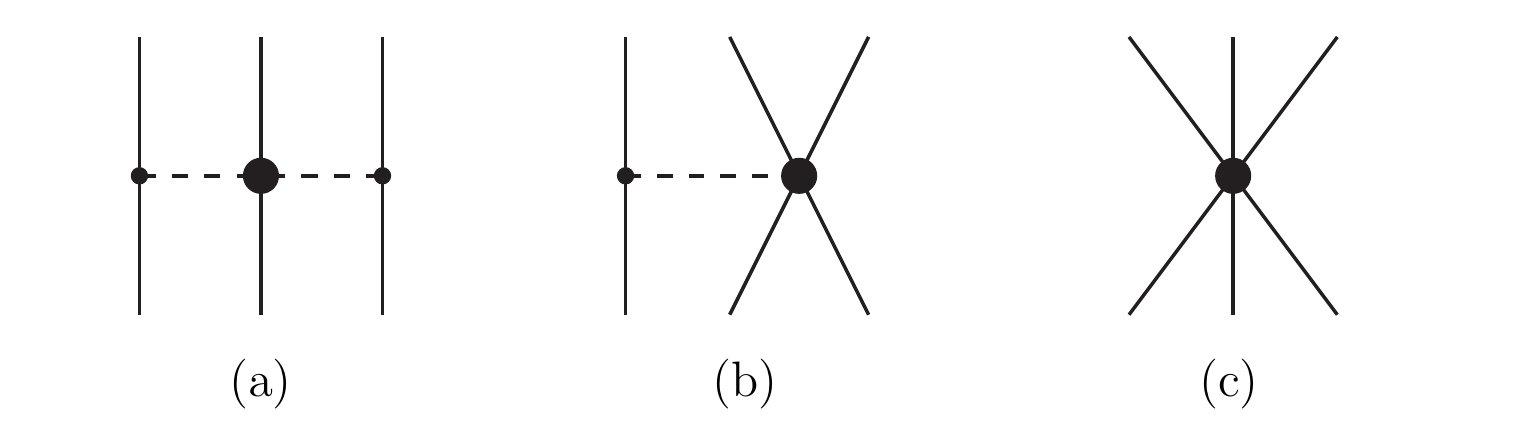}
\caption{Diagrams for the chiral three-nucleon interaction at N$^2$LO. In neutron matter, only 
diagram (a) contributes.} 
\label{3b}
\end{figure}

The leading three-nucleon force makes its appearance at third order in the chiral power 
counting and contains three contributions: the long-range 
two-pion-exchange part with $\pi\pi NN$ vertex proportional to the low-energy
constants $c_1,c_3,c_4$, the medium-range one-pion exchange diagram
proportional to the low-energy constant $c_D$, and finally the short-range contact 
term proportional to $c_E$. The corresponding diagrams are shown in 
Fig.~\ref{3b}, labeled as (a), (b), (c), respectively. Diagrams (b) and (c) vanish in 
neutron matter, while 
all three terms contribute in symmetric nuclear          
matter~\cite{holt10,hebeler10}.                                

Although efforts are in progress to incorporate 
potentially important N$^3$LO 3NF contributions 
\cite{ishikawa07,Ber08,Ber11} both in the neutron and nuclear 
equations of state and the fitting of the relevant low-energy constants, the ``N$^3$LO'' study reported in this paper 
is limited to the inclusion of the N$^2$LO three-body force together with the N$^3$LO 
two-body force, an approximation that is commonly used in the literature. The   
associated uncertainties for neutron matter have been investigated in Ref.~\cite{Sam+15}.                   

\begin{figure}[!t] 
\centering
\scalebox{0.65}{\includegraphics{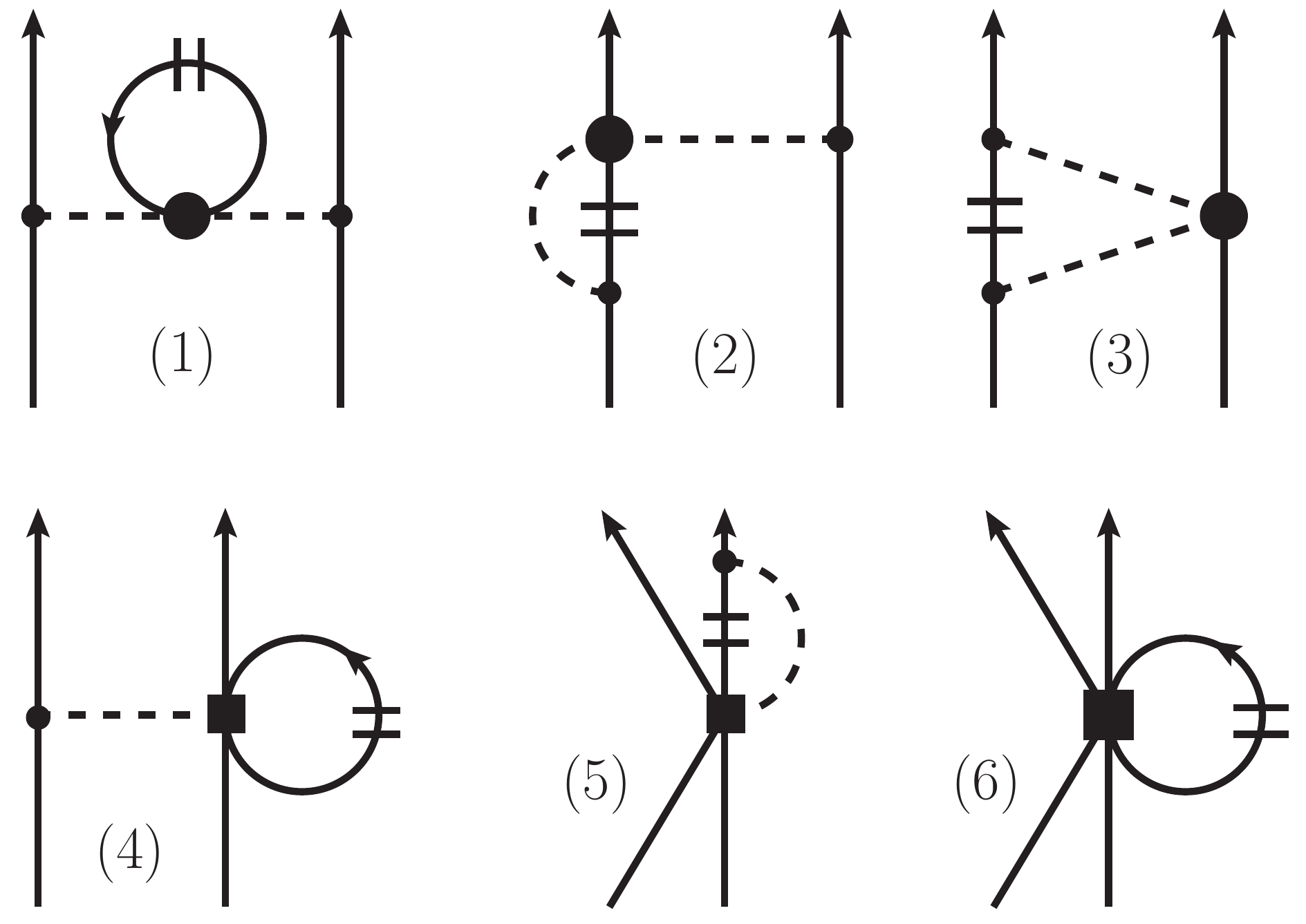}} 
\caption{Diagrams for the in-medium $NN$ interactions corresponding to $V_{NN}^{med,i}$ (i=1,...,6)
 given in the text. 
} 
\label{vnnmed}
\end{figure}

To facilitate the inclusion of 3NFs in the particle-particle ladder calculation, we employ the
density-dependent $NN$ interaction derived in Refs.~\cite{holt09,holt10} from
the N$^2$LO chiral three-body force. This effective interaction is obtained by
summing one particle line over the occupied states in the Fermi sea.            
Neglecting 
small contributions \cite{hebeler10} from terms depending on the center-of-mass 
momentum, the resulting $NN$ interaction can be expressed in analytical
form with operator structures identical to those of free-space $NN$ interactions, and
are therefore included on the same footing as two-body forces.
The small uncertainty associated with the use of these effective density-dependent 3NFs was discussed in Ref.~\cite{Sam+15}.

For the case of polarized isospin-asymmetric matter, the expressions from 
Ref.~\cite{holt10} are to be extended to include four different Fermi momenta, namely those
of upward(downward) polarized neutrons(protons), as described below.

Using the notation established above to indicate the Fermi
momenta 
of spin-up and spin-down neutrons or protons, 
the neutron and proton densities are                      
given by
$\rho_n=[(k_F^{nu})^3+(k_F^{nd})^3]/6\pi^2$ and 
$\rho_p=[(k_F^{pu})^3+(k_F^{pd})^3]/6\pi^2$.

Concerning kinematics, 
we consider elastic scattering process $N_1(\vec p\,)+N_2(-\vec p\,)\to N_1(\vec 
p+\vec q\,)+
N_2(-\vec p-\vec q\,)$ in the center-of-mass frame. 

Following the notation of Ref.~\cite{holt10}, we can distinguish between six effective density-dependent
NN interactions, denoted by diagram (1) to (6) in Fig.~2. They are: \\ 
The Pauli blocked pion-selfenergy (diagram (1)):
\begin{equation} V_{NN}^{\rm med,1}= {g_A^2 \over 2f_\pi^4}\,\vec \tau_1
\cdot \vec \tau_2 \,{\vec \sigma_1 \cdot\vec q \,\vec \sigma_2 \cdot \vec q
\over (m_\pi^2 + q^2)^2}\,(2c_1 m_\pi^2 +c_3 
q^2)(\rho_p+\rho_n)\,,\end{equation}

The Pauli blocked vertex correction (diagram (2)):
\begin{eqnarray} V_{NN}^{\rm med,2}&=& {g_A^2 \over 16\pi^2 f_\pi^4}\vec 
\tau_1
\cdot \vec \tau_2 \,  {\vec \sigma_1 \cdot \vec q \,\vec \sigma_2 \cdot 
\vec q
\over m_\pi^2 + q^2}\, \bigg\{-4c_1 m_\pi^2 
\Big[\Gamma_0^+(p)+\Gamma_1^+(p) \Big]
- (c_3+c_4)\nonumber \\ && \times \Big[q^2 \Big(\Gamma_0^+(p)+
2\Gamma_1^+(p)+\Gamma_3^+(p)\Big)+4\Gamma_2^+(p)\Big] + 4c_4 \Big[ 
2\pi^2(\rho_p+\rho_n)
-m_\pi^2\Gamma_0^+(p)\Big] \bigg\}\nonumber \\ && +{g_A^2 \over 32\pi^2 
f_\pi^4}
(\tau_1^3+\tau_2^3)\, {\vec \sigma_1 \cdot \vec q \,\vec \sigma_2 \cdot 
\vec q
\over m_\pi^2 + q^2}\, \bigg\{-4c_1 m_\pi^2 
\Big[\Gamma_0^-(p)+\Gamma_1^-(p) \Big]
+(c_4-c_3)\nonumber \\ && \times \Big[q^2 \Big(\Gamma_0^-(p)+2\Gamma_1^-(p)+
\Gamma_3^-(p)\Big)+ 4\Gamma_2^-(p)\Big] + 4c_4 \Big[ 2\pi^2(\rho_n-\rho_p)
+m_\pi^2\Gamma_0^-(p)\Big] \bigg\}\nonumber \\ && +{g_A^2 \over 16\pi^2 
f_\pi^4}
(\tau_1^3-\tau_2^3)\, i(\vec \sigma_1 -\vec \sigma_2) \cdot (\vec p \times
\vec q\,) {1 \over m_\pi^2+4p^2-q^2}\nonumber \\ && \times \bigg\{4c_1 
m_\pi^2
\Big[\Gamma_0^-(p)+\Gamma_1^-(p) \Big]+c_3(4p^2-q^2) \Big[\Gamma_0^-(p)+2
\Gamma_1^-(p)+\Gamma_3^-(p)\Big] \bigg\}\,,\end{eqnarray}
The last contribution, proportional to $(\tau_1^3-\tau_2^3)$, leads to
spin-singlet and spin-triplet mixing in the medium. It has been
Fierz-transformed to bring it into the form of the anti-symmetric 
spin-orbit
operator  $i(\vec \sigma_1 -\vec \sigma_2) \cdot (\vec p \times\vec q\,)$.
Terms which break rotational invariance in momentum-space due to the
spin-polarization of the nuclear medium in $z$-direction have been 
discarded.

Next, we give the expression for the Pauli blocked two-pion exchange (diagram (3)): 
\begin{eqnarray} V_{NN}^{\rm med,3} &=& {g_A^2 \over 32 
\pi^2f_\pi^4}\bigg\{  -12
c_1 m_\pi^2 \Big[2\Gamma_0^+(p)- (2m_\pi^2+q^2) G_0^+(p,q)\Big] 
\nonumber \\ &&
-3c_3 \Big[8\pi^2(\rho_p+\rho_n)-4(2m_\pi^2+q^2) \Gamma_0^+(p) 
-2q^2\Gamma_1^+(p)
+(2m_\pi^2+q^2)^2 G_0^+(p,q)\Big]\nonumber \\
&& +4c_4\,\vec \tau_1 \cdot  \vec \tau_2\, (\vec \sigma_1
\cdot \vec \sigma_2\, q^2 - \vec \sigma_1 \cdot  \vec q \,\vec\sigma_2\cdot
\vec q\,) G_2^+(p,q) \nonumber \\ && -(3c_3+c_4\vec \tau_1 \cdot \vec 
\tau_2 )\,
i ( \vec \sigma_1 +\vec \sigma_2 )\cdot(\vec q \times \vec 
p\,)\Big[2\Gamma_0^+(p)+
2\Gamma_1^+(p) - (2m_\pi^2+q^2)\nonumber \\ &&\times \Big(G_0^+(p,q)+2 
G_1^+(p,q)
\Big) \Big] -12 c_1 m_\pi^2\, i ( \vec \sigma_1 +\vec \sigma_2 ) 
\cdot(\vec q
\times \vec p\,)  \Big[G_0^+(p,q)+2 G_1^+(p,q)\Big] \nonumber \\ && + 
4c_4\,
\vec\tau_1 \cdot \vec \tau_2 \,\vec \sigma_1 \cdot (\vec q \times \vec 
p\,)\,
\vec\sigma_2 \cdot( \vec q \times \vec p \,) \Big[G_0^+(p,q) +
4G_1^+(p,q)+4G_3^+(p,q) \Big]\bigg\}\nonumber \\ &&
+{g_A^2 \over 64 \pi^2f_\pi^4}(\tau_1^3+\tau_2^3) \bigg\{  4 c_1 m_\pi^2 
\Big[
2\Gamma_0^-(p)- (2m_\pi^2+q^2) G_0^-(p,q)\Big] \nonumber \\ && 
+c_3\Big[8\pi^2
(\rho_p-\rho_n)-4(2m_\pi^2+q^2) \Gamma_0^-(p) 
-2q^2\Gamma_1^-(p)+(2m_\pi^2+q^2)^2
G_0^-(p,q)\Big]\nonumber \\&& -4c_4\, (\vec \sigma_1
\cdot \vec \sigma_2\, q^2 - \vec \sigma_1 \cdot  \vec q \,\vec\sigma_2\cdot
\vec q\,) G_2^-(p,q) \nonumber \\ && +(c_3+c_4)\,
i ( \vec \sigma_1 +\vec \sigma_2 )\cdot(\vec q \times \vec 
p\,)\Big[2\Gamma_0^-(p)+
2\Gamma_1^-(p) - (2m_\pi^2+q^2)\nonumber \\ &&\times \Big(G_0^-(p,q)+2 
G_1^-(p,q)
\Big) \Big] +4 c_1 m_\pi^2\, i ( \vec \sigma_1 +\vec \sigma_2 ) 
\cdot(\vec q
\times \vec p\,)  \Big[G_0^-(p,q)+2 G_1^-(p,q)\Big] \nonumber \\ && - 
4c_4\,
\vec \sigma_1 \cdot (\vec q \times \vec p\,)\, \vec\sigma_2 \cdot( \vec 
q \times
\vec p \,) \Big[G_0^-(p,q) +4G_1^-(p,q)+4G_3^-(p,q) \Big]\bigg\}
\,. \end{eqnarray}
The loop functions $\Gamma_j^\pm(p)$ and $G_j^\pm(p,q)$ with a 
superscript $+$ or $-$ are given by: 
\begin{equation} \Gamma_j^\pm(p) = {1\over 2}\Big[ 
\Gamma_j(p,k_{pu})+
\Gamma_j(p,k_{pd})\Big] \pm {1\over 2}\Big[ 
\Gamma_j(p,k_{nu})+
\Gamma_j(p,k_{nd})\Big]\,, \end{equation}
\begin{equation} G_j^\pm(p,q) = {1\over 2}\Big[ G_j(p,q,k_{pu})+
G_j(p,q,k_{pd})\Big] \pm {1\over 2}\Big[ G_j(p,q,k_{nu})+
G_j(p,q,k_{nd})\Big]\,, \end{equation}
where $\Gamma_j(p,k_f)$ and  $G(p,q,k_f)$ are defined in Eqs.~(13-16) and 
Eqs.~(18-22) of Ref.~\cite{holt10}. 

Now we present the 
contributions from the $1\pi$-exchange 3NF proportional to the low-energy constant $c_D$.  
The vertex correction to $1\pi$-exchange  linear in proton and neutron 
densities is (diagram (4)): 
\begin{equation} V_{NN}^{\rm med,4} = {g_A c_D \over 16f_\pi^4\Lambda_\chi}
\Big[-2\vec \tau_1 \cdot \vec \tau_2 (\rho_p+\rho_n)+ (\tau_1^3+ 
\tau_2^3)(\rho_p-
\rho_n)\Big] \,{\vec \sigma_1 \cdot\vec q \,\vec \sigma_2
\cdot \vec q  \over  m_\pi^2 + q^2} \,.\end{equation}

Pauli-blocking, diagram (5), contributes: 
\begin{eqnarray} V_{NN}^{\rm med,5}&=& {g_A c_D\over 32\pi^2 
f_\pi^4\Lambda_\chi}\bigg\{\vec
\tau_1 \cdot \vec \tau_2 \bigg[2 \vec \sigma_1 \cdot \vec 
\sigma_2\,\Gamma_2^+(p)
  +\bigg(\vec \sigma_1 \cdot \vec \sigma_2 \Big( 2p^2-{q^2\over 2}\Big) 
+\vec
\sigma_1 \cdot \vec q\,\vec\sigma_2  \cdot \vec q\,\nonumber \\ && \times
\Big(1-{2p^2\over q^2}\Big) -{2\over q^2}\,\vec\sigma_1 \cdot (\vec q 
\times
\vec p\,)\,\vec\sigma_2 \cdot(\vec q\times \vec p \,)\bigg)
\Big[\Gamma_0^+(p)+2\Gamma_1^+(p)+\Gamma_3^+(p)\Big] \bigg]\nonumber \\ &&
+12\pi^2(\rho_p+\rho_n)-6m_\pi^2 \Gamma_0^+(p)\bigg\}\nonumber \\ && 
+{g_A c_D\over
64\pi^2 f_\pi^4\Lambda_\chi}(\tau_1^3+\tau_2^3)\bigg\{2\vec\sigma_1 
\cdot \vec \sigma_2\,
\Gamma_2^-(p) +\bigg[\vec \sigma_1 \cdot \vec \sigma_2 \Big( 
2p^2-{q^2\over 2}\Big)
+\vec \sigma_1 \cdot \vec q\,\vec\sigma_2  \cdot \vec q\,\nonumber \\ && 
\times
\Big(1-{2p^2\over q^2}\Big) -{2\over q^2}\,\vec\sigma_1 \cdot (\vec q 
\times
\vec p\,)\,\vec\sigma_2 \cdot(\vec q\times \vec p \,)\bigg]
\Big[\Gamma_0^-(p)+2\Gamma_1^-(p)+\Gamma_3^-(p)\Big]\nonumber \\ &&
+4\pi^2(\rho_n-\rho_p)+2m_\pi^2 \Gamma_0^-(p)\bigg\}\,. \end{eqnarray}

The contribution from the contact 3NF proportional to the low-energy constant $c_E$ is (diagram (6)): 
\begin{equation}  V_{NN}^{\rm med,6} ={3c_E \over4 
f_\pi^4\Lambda_\chi}\Big[-2
(\rho_p+\rho_n) +(\rho_p-\rho_n)(\tau_1^3+\tau_2^3)\Big] \,.\end{equation}

Partial wave matrix elements with  $J\geq 1$ of the antisymmetric 
spin-orbit term, which occur in Eq.~(14), mix
spin-singlet and spin-triplet states and these can be calculated for 
on-shell kinematics in
the center-of-mass frame as:
\begin{eqnarray} && \langle J0J|  i(\vec \sigma_1 \!-\!\vec 
\sigma_2)\!\cdot\! (\vec p
\times \vec q\,)F(q^2)|J1J\rangle =  \langle J1J|  i(\vec 
\sigma_1\!-\!\vec \sigma_2)\!
\cdot\!(\vec p \times\vec q\,)F(p^2,q^2)|J0J\rangle \nonumber \\ &&= 
{\sqrt{J(J+1)} \over
2J+1} \int_{-1}^1 \!dz \, p^2 F(p^2,2p^2(1-z)) \Big[ P_{J-1}(z)- 
P_{J+1}(z)\Big] \,.
\end{eqnarray}
However, because of the small size of this contribution, particularly for small proton fractions,
we neglect this term in the present calculations.

\section{Results and discussion}                                                                  
\label{res} 

\begin{figure}[!t] 
\centering 
\vspace*{-2.0cm}
\hspace*{-1.0cm}
\scalebox{0.65}{\includegraphics{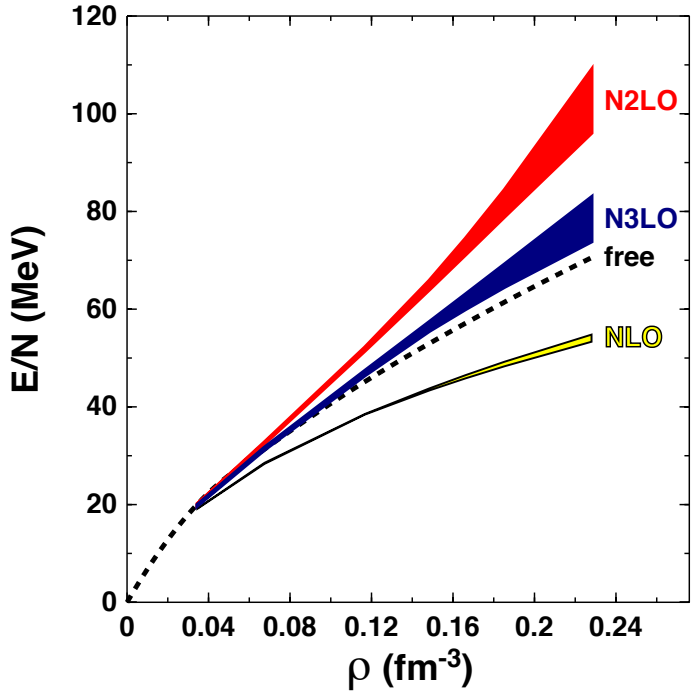}} 
\vspace*{-4.2cm}
\caption{(color online)                                        
Energy per neutron in fully polarized neutron matter as a function 
of density. The yellow and red bands represent the uncertainities due to cutoff variations obtained
in the complete calculations at NLO and N$^2$LO, respectively. The blue band is the result of the 
same cutoff variations applied to our exploratory N$^3$LO calculation, see text for details. 
The dotted curve shows the energy of the free Fermi gas.
} 
\label{pnm}
\end{figure}

We show in Fig.~\ref{pnm} the energy per particle in fully polarized neutron matter
as a function of density. The yellow and red bands represent the predictions of 
complete calculations at second and third 
order, respectively, of chiral effective field theory, while the blue band shows the predictions obtained with the exploratory N$^3$LO calculation as described
above. 
For each band, the width is obtained by changing the cutoff between 450 MeV and 600 MeV.

At N$^2$LO and N$^3$LO, 
cutoff dependence is generally moderate up to saturation density. At NLO, the cutoff dependence is 
practically negligible throughout.
In unpolarized
neutron matter, on the other hand, the largest cutoff dependence was seen at NLO~\cite{Sam+15}. This suggests that, in unpolarized NM,
the larger cutoff sensitivity at NLO is mostly due to singlet states, particularly $^1S_0$, which are absent from the polarized system.
At the same time, 3NFs do not appear at NLO, implying that most of the cutoff dependence in polarized NM
at N$^2$LO and N$^3$LO is caused by the 3NF contributions. 

Clearly, the variations associated with changing the cutoff are not a good 
indicator of the uncertainty at a given order of chiral effective field theory, 
as the results from one order to the other do not overlap.
Furthermore, the predictions do not show a good convergence pattern, although 
some indication of slow convergence can be seen when moving from N$^2$LO to our N$^3$LO calculation.

As can be concluded from Table~\ref{tab2}, the predictions from 
the N$^3$LO calculation are close to the free Fermi gas energy, at least up
to saturation densities. Our results with the N$^3$LO ($\Lambda$=500 MeV) potential are in good agreement 
 with those from Ref.~\cite{krueg14} using the same potential as well as three- and four-nucleon forces at 
 N$^3$LO.                                                                   
With regard to the similarity with the free Fermi gas, it is interesting to include some additional considerations.
As mentioned in the Introduction, many-fermion systems with large scattering lengths offer the 
opportunity to model low-density neutron matter. In the unitary limit (that is, when the system can support a 
bound state at zero energy), the scattering length approaches infinity. The system then becomes
scale-independent and the ground-state energy is
determined by a single universal parameter, known as the Bertsch parameter, $\xi$. 
The latter is defined as the ratio of the energy per particle of the unitary gas to that of the free 
Fermi gas. In Ref.~\cite{Kais12}, using a simple ansatz for the interaction,
 it is shown that $\xi$ increases from approximately 0.5 to 
1.0 as the spin asymmetry of neutron matter, $\beta_n$, is increased from 0 (unpolarized) to 1 (fully polarized).

\begin{figure}[!t] 
\centering 
\vspace*{-2.0cm}
\hspace*{-1.0cm}
\scalebox{0.65}{\includegraphics{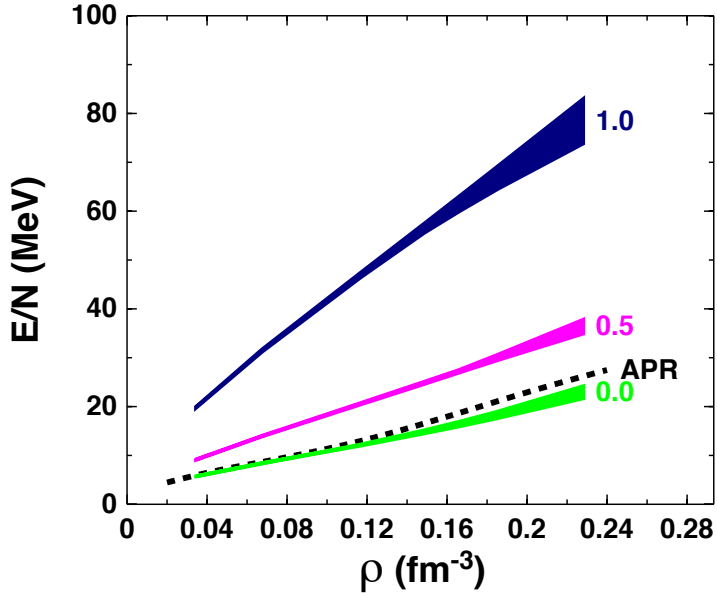}} 
\vspace*{-4.2cm}
\caption{(color online)                                        
Energy per neutron in pure neutron matter as a function of density at N$^3$LO. From lowest to highest curve:
unpolarized NM; partially polarized NM, with $\beta_n$=0.5; fully polarized NM ($\beta_n$=1). 
The width of each band shows the uncertainty from varying the cutoff between 
450 and 600 MeV. The black dotted line shows the predictions for the equation of state of unpolarized
neutron matter from Ref.~\cite{APR}. 
} 
\label{beta}
\end{figure}

In Fig.~\ref{beta}, for our N$^3$LO calculation, we compare predictions (along with their cutoff
variations) of the energy per neutron in: unpolarized NM (green band), partially polarized NM (pink band), and
fully polarized NM (blue band). 
For the partially polarized case, the value of $\beta_n$ (see Eq.~(2)) is equal to 0.5, corresponding to 75\% 
of the neutrons being polarized in one direction and 25\% in the opposite direction, see Eqs.~(3-4).
Clearly, a lesser degree of spin asymmetry (as compared to the ferromagnetic case) yields considerably
less repulsion. 
There is definitely no sign of a phase transition, particularly to a ferromagnetic state, nor an indication that 
such transition may occurr at higher densities. This is consistent with what we observed earlier~\cite{Sam11} 
with meson-theoretic interactions. 

As a baseline comparison, we also include, for the unpolarized case, predictions based on a different
approach, shown by the black dotted line in Fig.~\ref{beta}.
These are taken from Ref.~\cite{APR} and are based on the Argonne $v_{18}$ 
two-nucleon interaction 
plus the Urbana IX three body-force, using variational methods. The predictions are overall in reasonable agreement with our green band, although those from 
Ref.~\cite{APR} show more repulsion as compared to the softer chiral interactions.

\begin{figure}[!t] 
\centering 
\vspace*{-2.0cm}
\hspace*{-1.0cm}
\scalebox{0.65}{\includegraphics{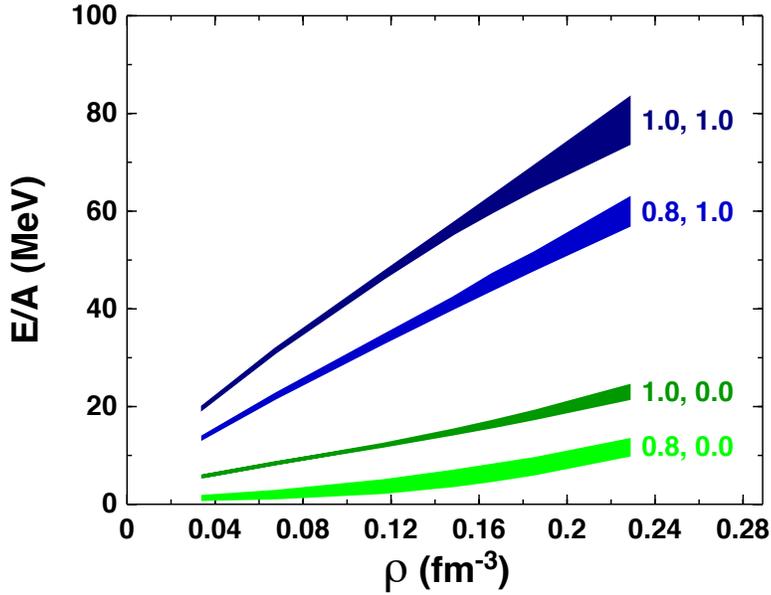}} 
\vspace*{-4.2cm}
\caption{(color online)                                        
Energy per nucleon in neutron-rich matter 
as a function of density at N$^3$LO and different conditions of isospin and spin polarization.
The (brighter blue) band labeled as ``0.8, 1.0" displays the results for neutron-rich matter with a 
proton fraction equal to 10\% ($\alpha$=0.8) and fully polarized neutron ($\beta_n$=1.0).                      
The (brighter green) band labeled as ``0.8, 0.0" refers to neutron-rich matter with the same proton fraction 
and no polarization ($\beta_n$=0.0). The protons are unpolarized. 
For comparison, we also include the bands (darker blue and darker green) already shown in the previous figure, 
which refer to pure neutron matter ($\alpha$=1) with fully polarized ($\beta_n$=1) or 
unpolarized ($\beta_n$=0) neutrons.   
The bands are obtained varying the cutoff between 450 and 600 MeV.
} 
\label{alpha}
\end{figure}

Most typically, models which do predict spin instability of neutron matter find the phase transition to 
occurr at densities a few times normal density. Such high densities are outside the domain of chiral perturbation
theory. 
With some effective forces, though, it was found~\cite{pol19} that a small fraction of protons can significantly reduce the 
onset of the threshold density for a phase transition to a spin-polarized state of neutron-rich matter.
We explored this scenario by adding a small fraction of protons to fully polarized or unpolarized neutrons.
From Eqs.~\ref{rho}-\ref{rhopnud}, a proton fraction of 10\% is obtained with $\alpha$=0.8.
The results are displayed in 
Fig.~\ref{alpha}, where a crossing of the bands labeled with ``0.8, 1.0" and ``0.8, 0.0", respectively,
would indicate a phase transition. Thus we conclude that such transition is not predicted with chiral 
forces. By extrapolation, a transition to a polarized state would also appear very unlikely at higher densities. 

\begin{table}                
\centering
\begin{tabular}{|c||c|c|}
\hline
Density (fm$^{-3}$) & $\Lambda$ (MeV) & $E_{FFG}/E$ \\
\hline     
  0.15 & 450 & 0.95 \\
       & 500 & 0.92  \\
       & 600 & 0.95  \\
  0.17 & 450 & 0.95  \\
       & 500 & 0.91  \\
       & 600 & 0.93  \\
\hline
\end{tabular}
\caption                                                    
{Ratio of the energy per particle of a free Fermi gas 
to the energy per particle of polarized neutron matter 
around saturation density at N$^3$LO (as described in the text)
and for different values of the cutoff.} 
\label{tab2}
\end{table}

\section{Conclusions and outlook}                                                                  

We have calculated the equation of state of (fully and partially) polarized neutron-rich matter.
We performed complete calculations at second and third order of chiral effective field
theory and calculations employing the N$^3$LO 2NF plus the leading 3NF.
Results with both spin and isospin asymmetries are presented for the first time with chiral forces.

In all calculations, the 
cutoff dependence is moderate and definitely underestimates the uncertainty of each order.  
Concerning the latter, we do not see a satisfactory convergence pattern. The missing
3NFs are most likely not the main cause of uncertainty at N$^3$LO, since               
Ref.~\cite{krueg14} has demonstrated that large cancelations take place between 
the 2$\pi$-exchange 3NF and the $\pi$-ring 3NF at N$^3$LO, while other 3NF contributions are very small (about
0.1-0.2 MeV). 
Clearly a calculation at 
N$^4$LO is absolutely necessary to get a realistic indication of the EFT error at 
N$^3$LO. Such effort is in progress.                         
If such calculation displays a reasonable convergence pattern, it will be strong evidence that 
polarized neutron matter, indeed, behaves nearly like a free Fermi gas, at least up to normal densities.

In our N$^3$LO calculation,                                         
the energies of the unpolarized system at normal density are close to 16 MeV for all cutoffs, 
whereas those in the polarized case are approximately 60 MeV. Thus, even in the presence
of the large uncertainties discussed above, 
a phase transition to a ferromagnetic state can be excluded.
This conclusion remaind valid in the presence of a small proton fraction. 

\section*{Acknowledgments}
This work was supported in part by 
the U.S. Department of Energy Office of Science, Office of Basic Energy Science, under Award No. DE-FG02-03ER41270 
(F.S. and R.M.), and by DFG and NSFC (CRC110) (N.K.).

\end{document}